\documentclass[10pt,conference,letterpaper]{IEEEtran}
\IEEEoverridecommandlockouts

\usepackage{cite}
\usepackage{amsmath,amssymb,amsfonts}
\usepackage{graphicx}
\usepackage{textcomp}
\usepackage{xcolor}
\usepackage{booktabs}
\usepackage{multirow}
\usepackage{tabularx}
\usepackage{hyperref}

\def\BibTeX{{\rm B\kern-.05em{\sc i\kern-.025em b}\kern-.08em
    T\kern-.1667em\lower.7ex\hbox{E}\kern-.125emX}}

\begin{document}

\title{Rank-Refined Quantum-Behaved Particle Swarm Optimization for Quantum Molecular Generation}

\author{
\IEEEauthorblockN{
    Sing-Yun Wu\IEEEauthorrefmark{1},
    Sheng Yun Wu\IEEEauthorrefmark{2},
    I-Min Chiang\IEEEauthorrefmark{1},
    and Tai-Yue Li\IEEEauthorrefmark{3}
}
\IEEEauthorblockA{\IEEEauthorrefmark{1} Tzu Chi Senior High School Affiliated with Tzu Chi University, Hualien, Taiwan}
\IEEEauthorblockA{\IEEEauthorrefmark{2} Department of Physics, National Dong Hwa University, Hualien, Taiwan}
\IEEEauthorblockA{\IEEEauthorrefmark{3} National Center for High-performance Computing, National Institutes of Applied Research, Hsinchu, Taiwan}
\IEEEauthorblockA{
Emails: sirius823935@gmail.com, sywu@mail.ndhu.edu.tw, emin919@gmail.com, tim312508@gmail.com
}
}

\maketitle

\begin{abstract}
This work proposes Rank-Refined Quantum-Behaved Particle Swarm Optimization (RR-QPSO) for high-dimensional parameter search in Quantum Molecular Generation (QMG). RR-QPSO targets the optimization bottleneck caused by expensive objective evaluations, where each candidate parameter vector requires stochastic circuit sampling, bitstring decoding, and molecular evaluation. The method serves as a population-based alternative to Bayesian optimization (BO), combining Sobol-based initialization, a rank-refined mean-best update, and fitness-guided refinement based on validity and uniqueness. Experiments on the 9-heavy-atom QMG benchmark use a 134-parameter, 20-qubit dynamic quantum circuit simulated with CUDA-Q, with particle evaluations parallelized across 8 NVIDIA V100 GPUs. With $M=64$ particles and $T=150$ iterations, RR-QPSO reaches a validity--uniqueness product of $V \times U = 0.930$. Increasing the swarm size to $M=128$ further improves the product to $V \times U = 0.942$, compared with $0.902$ for a BO baseline. A multi-objective extension targeting $\mathrm{HBA}=4$ and $\mathrm{HBD}=3$ also shows that RR-QPSO can guide molecular properties while retaining stronger validity--uniqueness performance than BO in the same scalarized setting. These results demonstrate that optimizer-level design can improve QMG performance without modifying the chemistry-inspired circuit or molecular decoding pipeline.
\end{abstract}


\section{Introduction}
De novo molecular generation aims to sample novel and chemically valid molecules from an extremely large chemical space, whose size is estimated to exceed $10^{33}$ drug-like structures~\cite{polishchuk2013estimation}. Classical machine learning has driven much of the recent progress in this area. Variational autoencoders learn continuous molecular representations that can be decoded into new compounds~\cite{gomezbombarelli2018automatic,jin2018junction}, recurrent models generate Simplified Molecular Input Line Entry System (SMILES) strings token by token~\cite{segler2018generating}, and graph-based generative models such as MolGAN operate directly on molecular graphs~\cite{decao2018molgan}. Although these approaches have shown strong generative performance, they often depend on large curated datasets and may suffer from mode collapse or off-distribution generation. These limitations motivate alternative generative models that can represent molecular structure through compact and sample-based mechanisms.

Quantum machine learning (QML) provides one such direction through parameterized quantum circuits~\cite{biamonte2017quantum,cerezo2021variational}. By preparing and measuring quantum states, these circuits can represent probability distributions over high-dimensional bitstrings using a compact set of trainable gates, making them attractive for resource-constrained and sampling-oriented learning problems. Recent QML studies have explored variational quantum circuits in hybrid neural modules, quantum kernels, and measurement-probability models~\cite{rebentrost2014quantum,havlivcek2019supervised,tai2022quantum,an2025quantum,hsu2025qae,rai2026hybrid,wang2026mpmqir}. At the same time, tensor-network and GPU-accelerated simulation have enabled validation of larger QML circuits beyond direct state-vector simulation~\cite{chen2024validating,sam2026dual,sam2026grover}. These advances make chemically structured generation a natural application domain for sampling-based quantum models.

Quantum-assisted molecular generation has been explored through quantum generative adversarial networks, quantum circuit Born machines (QCBMs), and QML-based drug-discovery models~\cite{dallaire2018quantum,liu2018differentiable,li2021drug,thomas2025qca}. Quantum molecular generation (QMG) follows a chemistry-inspired route by encoding atom and bond decisions into a dynamic quantum circuit~\cite{chen2025chemistry}. The circuit generates molecules sequentially through mid-circuit measurements and classically controlled operations, producing molecular graphs directly from quantum samples. QMG is appealing because it uses a compact qubit representation and does not require a large molecular training corpus for unconditional generation. The original QMG study used Bayesian optimization (BO), a standard approach for expensive black-box optimization~\cite{snoek2012practical}, to tune the circuit parameters according to validity and uniqueness. However, QMG parameter search remains challenging because the circuit parameter space grows with the molecular generation setting, and each candidate vector must be evaluated through thousands of circuit samples, bitstring decoding, and molecular validity checks. This makes QMG parameter optimization high-dimensional, stochastic, and expensive, where optimizer design directly affects molecular-generation quality.

In this work, we address the optimizer bottleneck in QMG with Rank-Refined Quantum-Behaved Particle Swarm Optimization (RR-QPSO), replacing the BO-based parameter search with a population-based optimizer. The contributions are threefold. First, we formulate QMG parameter search as a stochastic black-box optimization problem whose candidate evaluations are independent and therefore naturally parallelizable. Second, we adapt Quantum-Behaved Particle Swarm Optimization (QPSO) to the QMG setting through Sobol-based initialization for broad initial coverage, a rank-refined mean-best update for stronger swarm-level search guidance, and fitness-guided refinement to balance validity and uniqueness. Third, we evaluate RR-QPSO on the 9-heavy-atom QMG benchmark and further demonstrate its extension to multi-objective molecular generation targeting hydrogen-bond acceptor (HBA) and hydrogen-bond donor (HBD) counts. Together, these contributions show that RR-QPSO provides a parallelizable optimizer for QMG across both molecular-quality and property-guided objectives.

\section{Quantum Molecular Generation Preliminaries}
\label{sec:qmg}
QMG formulates molecular generation as a parameterized quantum sampling problem in which a chemistry-inspired dynamic quantum circuit encodes sequential atom and bond decisions~\cite{chen2025chemistry}. For molecules with up to $N$ heavy atoms, the trainable circuit is controlled by $\boldsymbol{\theta}\in\mathbb{R}^{D}$, where
\begin{equation}
D = 8 + \frac{3(N-2)(N+3)}{2}.
\end{equation}
In this work, we follow the 9-heavy-atom setting, which gives $D=134$. The optimization task is therefore to tune $\boldsymbol{\theta}$ so that the induced sampling distribution yields high-quality decoded molecules.

The QMG circuit generates candidate molecules through an atom-then-bond construction process. At each growth step, an atom-generation part samples the graph-growth status and, when growth continues, the type of the newly introduced atom. Conditional bond-generation parts then sample the bond types between the new atom and previously generated atoms. Mid-circuit measurements provide the classical information needed to activate later operations only when the corresponding atom or bond decision is relevant, giving the circuit a dynamic structure that follows molecular graph construction. The resulting bitstrings are decoded into molecular graphs, and the decoded structures are checked for chemical validity, including valence consistency and molecular connectivity, using the same evaluation criteria as the original QMG study. Thus, the quality of a circuit setting is observed only after the complete sampling, decoding, and molecular evaluation pipeline.

Each circuit setting is evaluated using validity and uniqueness. Given $N_{\mathrm{sample}}$ generated samples, let $N_{\mathrm{valid}}$ denote the number of chemically valid molecules and $N_{\mathrm{unique}}$ denote the number of unique molecules among the valid set. The two metrics are
\begin{equation}
V = \frac{N_{\mathrm{valid}}}{N_{\mathrm{sample}}},
\qquad
U = \frac{N_{\mathrm{unique}}}{N_{\mathrm{valid}}}.
\end{equation}
The primary fitness objective is the product
\begin{equation}
F(\boldsymbol{\theta}) = V(\boldsymbol{\theta}) \times U(\boldsymbol{\theta}).
\label{eq:fitness}
\end{equation}
This objective rewards configurations that generate valid molecules while preserving diversity. The resulting problem is a stochastic black-box optimization problem:
\begin{equation}
\boldsymbol{\theta}^{*}
=
\arg\max_{\boldsymbol{\theta}\in\Omega}
F(\boldsymbol{\theta}),
\end{equation}
where $\Omega$ denotes the bounded parameter domain. Since $F(\boldsymbol{\theta})$ is estimated from finite circuit samples, the objective is noisy and expensive to evaluate. The same sampling, decoding, and scoring pipeline is used for the re-run BO baseline and the QPSO-based optimizers in Section~\ref{sec:results}. The independence of candidate evaluations motivates a population-based parallel search strategy.

\section{Method}
\label{sec:method}
\subsection{Optimization Framework}
\begin{figure*}[!t]
  \centering
  \includegraphics[width=0.98\textwidth]{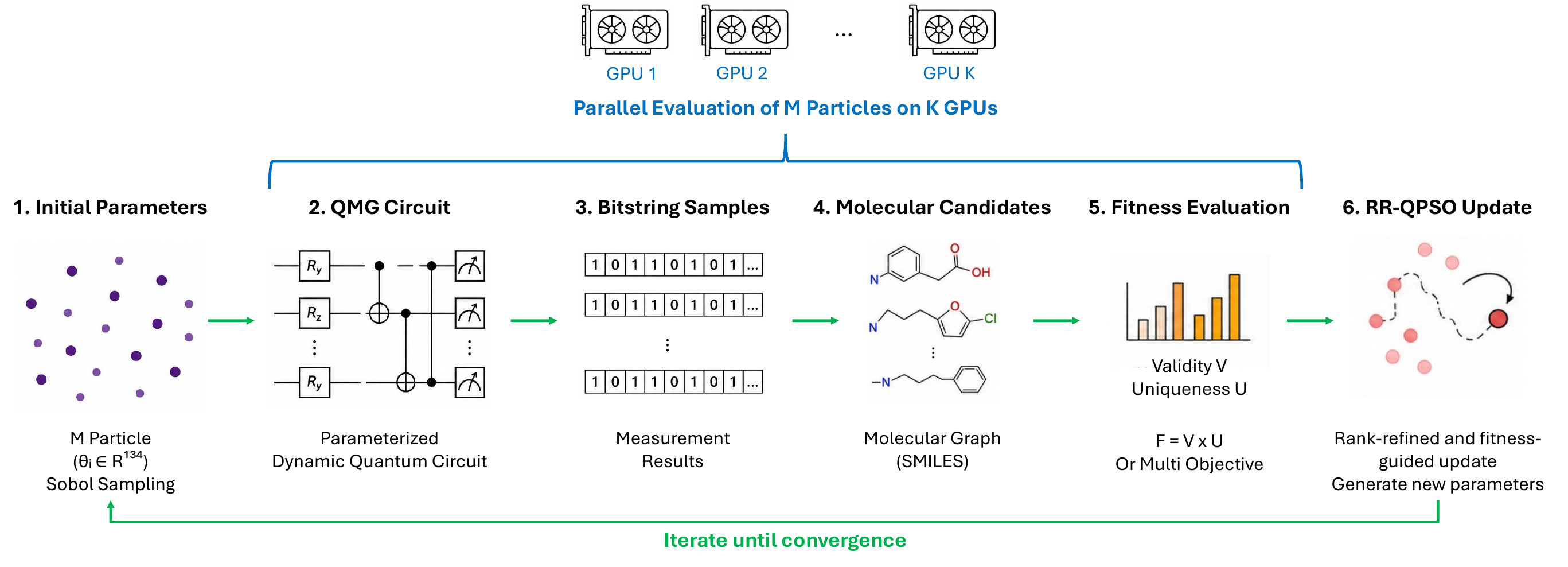}
  \caption{RR-QPSO workflow for QMG parameter optimization. Sobol sampling
  initializes a swarm of $M$ candidate parameter vectors, which are evaluated
  through QMG circuit sampling, bitstring decoding, molecular graph generation,
  and fitness computation. The independent particle evaluations are distributed
  across $K$ GPUs, and the resulting scores are used to update the RR-QPSO
  swarm.}
  \label{fig:workflow}
\end{figure*}

Figure~\ref{fig:workflow} summarizes the proposed RR-QPSO framework. The workflow consists of Sobol-based swarm initialization, parallel QMG evaluation, and RR-QPSO parameter update. The QMG evaluation stage includes circuit sampling, bitstring decoding, molecular candidate generation, and fitness computation. At iteration $t$, the optimizer maintains a swarm of $M$ particles,
\begin{equation}
\mathcal{X}^{t} = \{\boldsymbol{x}_{1}^{t}, \boldsymbol{x}_{2}^{t}, \ldots, \boldsymbol{x}_{M}^{t}\},
\end{equation}
where each particle $\boldsymbol{x}_{i}^{t}\in\mathbb{R}^{D}$ represents one candidate QMG parameter vector, with $D=134$ in the 9-heavy-atom benchmark. Each particle is evaluated by the QMG pipeline in Section~\ref{sec:qmg}: the dynamic circuit is sampled, the measured bitstrings are decoded into molecular graphs, and the resulting molecules are scored using Eq.~\eqref{eq:fitness}. The resulting scores update the personal best, global best, and swarm-level search direction before the next optimization round.

Unlike BO, which builds a surrogate model over the search space, RR-QPSO directly evolves a population of parameter vectors. This population-based design is well matched to QMG because candidate evaluations are expensive but mutually independent.

\subsection{Swarm Initialization}
The initial swarm influences how effectively the optimizer covers the $D$-dimensional parameter space. Consistent with the first block of Fig.~\ref{fig:workflow}, we use Owen-scrambled Sobol sampling~\cite{sobol1967distribution,owen1995randomly} to generate a reproducible low-discrepancy initial population. Each Sobol point $\boldsymbol{s}_i\in[0,1]^D$ is mapped to the bounded QMG parameter domain by
\begin{equation}
\boldsymbol{x}_{i}^{0}
=
\boldsymbol{\ell}
+
\boldsymbol{s}_{i}\odot(\boldsymbol{u}-\boldsymbol{\ell}),
\end{equation}
where $\boldsymbol{\ell}$ and $\boldsymbol{u}$ are the lower and upper bounds. This provides a space-filling starting population before reliable fitness information is available.

Although initialization alone does not guarantee better final performance, it provides a controlled and reproducible starting population for the subsequent RR-QPSO updates.

\subsection{RR-QPSO Parameter Update}
QPSO updates particles using a probabilistic position update rather than the velocity update used in classical PSO~\cite{sun2004particle,sun2012quantum}. Each particle maintains a personal best position $\boldsymbol{p}_{i}^{t}$, and the swarm maintains a global best position $\boldsymbol{g}^{t}$. The local attractor of particle $i$ is
\begin{equation}
\boldsymbol{a}_{i}^{t}
=
\boldsymbol{\phi}_{i}^{t}\odot\boldsymbol{p}_{i}^{t}
+
(1-\boldsymbol{\phi}_{i}^{t})\odot\boldsymbol{g}^{t},
\end{equation}
where $\boldsymbol{\phi}_{i}^{t}\sim U(0,1)^D$. The particle update is
\begin{equation}
\boldsymbol{x}_{i}^{t+1}
=
\boldsymbol{a}_{i}^{t}
+
\boldsymbol{b}_{i}^{t}
\odot
\alpha_{t}
\left|
\boldsymbol{m}^{t}-\boldsymbol{x}_{i}^{t}
\right|
\odot
\ln\left(\frac{1}{\boldsymbol{r}_{i}^{t}}\right),
\label{eq:qpso}
\end{equation}
where $\boldsymbol{r}_{i}^{t}\sim U(0,1)^D$, each component of $\boldsymbol{b}_{i}^{t}\in\{-1,+1\}^{D}$ is sampled uniformly to determine the update direction, $\alpha_t$ is a contraction--expansion coefficient, and $\boldsymbol{m}^{t}$ is the swarm mean-best position. In our implementation, $\alpha_t$ is bounded by $\alpha_{\max}=1.2$ and $\alpha_{\min}=0.3$ to support broad exploration early while limiting step sizes later in the search.

In standard QPSO, $\boldsymbol{m}^{t}$ is computed as the average of all personal best positions. We replace this average with a rank-refined mean-best update. After sorting personal best positions by fitness, let $\boldsymbol{p}_{(k)}^{t}$ denote the $k$-th best personal best and $\boldsymbol{p}_{(M-k+1)}^{t}$ denote the corresponding low-fitness personal best. The proposed mean-best position is
\begin{equation}
\boldsymbol{m}_{\mathrm{RR}}^{t}
=
\frac{1}{M}\sum_{i=1}^{M}\boldsymbol{p}_{i}^{t}
+
\rho_t
\sum_{k=1}^{\lfloor M/2 \rfloor}
\frac{
\boldsymbol{p}_{(k)}^{t}
-
\boldsymbol{p}_{(M-k+1)}^{t}
}{k},
\label{eq:rr_mbest}
\end{equation}
where $\rho_t$ controls the strength of the rank-based correction. In our experiments, we use a fixed correction strength, $\rho_t=\rho=0.015$, for all iterations. This update shifts the swarm attractor toward directions that separate high-fitness and low-fitness regions, providing a stronger population-level search signal than simple averaging.

\subsection{Fitness-Guided Refinement}
The product objective in Eq.~\eqref{eq:fitness} can hide different failure modes. Two parameter vectors may have similar $V\times U$ values while one generates many valid but repetitive molecules and the other generates diverse but mostly invalid molecules. To expose these complementary search directions, we track validity-oriented and uniqueness-oriented elite solutions in addition to the rank-refined mean-best vector.

We form a fitness-guided attractor
\begin{equation}
\tilde{\boldsymbol{m}}^{t}
=
\frac{
w_{\mathrm{RR}}\boldsymbol{m}_{\mathrm{RR}}^{t}
+
\mathbb{I}_{V}^{t}w_V\boldsymbol{x}_{V}^{t}
+
\mathbb{I}_{U}^{t}w_U\boldsymbol{x}_{U}^{t}
}{
w_{\mathrm{RR}}+\mathbb{I}_{V}^{t}w_V+\mathbb{I}_{U}^{t}w_U
}.
\label{eq:vu_mbest}
\end{equation}
Here $\boldsymbol{x}_{V}^{t}$ and $\boldsymbol{x}_{U}^{t}$ denote the best validity-oriented and uniqueness-oriented elite solutions observed up to iteration $t$. The indicators activate an elite only when its complementary metric is acceptable:
\begin{equation}
\mathbb{I}_{V}^{t}
=
\mathbf{1}\!\left[
U(\boldsymbol{x}_{V}^{t}) \ge \tau_U
\right],
\qquad
\mathbb{I}_{U}^{t}
=
\mathbf{1}\!\left[
V(\boldsymbol{x}_{U}^{t}) \ge \tau_V
\right].
\end{equation}
This prevents the update from following a high-validity solution with severe mode collapse or a high-uniqueness solution with poor validity. In our experiments, we set $\tau_U=\tau_V=0.5$, $w_{\mathrm{RR}}=0.70$, and $w_V=w_U=0.15$, so the update remains primarily driven by the RR mean-best while retaining auxiliary guidance from validity- and uniqueness-oriented elites. The guided attractor $\tilde{\boldsymbol{m}}^{t}$ replaces $\boldsymbol{m}^{t}$ in Eq.~\eqref{eq:qpso}. After each update, all particles are clipped to the valid parameter range.

\subsection{Multi-GPU Parallel Evaluation}
The QMG evaluation stage is naturally parallelizable because particle scores are independent once the candidate parameter vectors are generated. Consistent with Fig.~\ref{fig:workflow}, at each iteration the $M$ candidate parameter vectors are distributed across $K$ GPUs. Each GPU worker evaluates one or more particles by running QMG circuit sampling, molecular decoding, and fitness computation. The optimizer then collects the validity, uniqueness, and fitness scores before performing the RR-QPSO update.

This workflow is important because QMG evaluation dominates the runtime. In our implementation, particle evaluations are assigned to isolated worker processes, which improves robustness during repeated CUDA-Q simulations. The multi-GPU evaluation design enables larger particle counts, such as $M=128$, within the same optimization workflow.

\section{Results}
\label{sec:results}
All experiments are conducted on the 9-heavy-atom QMG benchmark using the 134-parameter, 20-qubit dynamic circuit described in Section~\ref{sec:qmg}. Unless otherwise stated, each candidate parameter vector is evaluated with 5000 shots, and the primary metric is the validity--uniqueness product $V\times U$. The first two experiments use the unconditional objective in Eq.~\eqref{eq:fitness}, whereas the final experiment uses the scalarized target-property objective for multi-objective optimization. Quantum circuit simulations are performed with CUDA-Q 0.7.1 using the cuStateVec backend on NVIDIA V100 GPUs.

\subsection{Performance Comparison}
The first experiment compares four optimizer configurations using the same validity--uniqueness metric: BO, QPSO without Sobol initialization, QPSO with Sobol initialization, and RR-QPSO. The BO baseline is re-run under the same implementation and evaluation protocol rather than taken directly from the literature. The QPSO-based runs use $M=64$ particles, $T=150$ iterations, and 5000-shot molecular evaluation. Because all methods use the same QMG circuit and scoring rule, the comparison focuses on the optimizer.

\begin{figure}[!t]
  \centering
  \includegraphics[width=0.98\columnwidth]{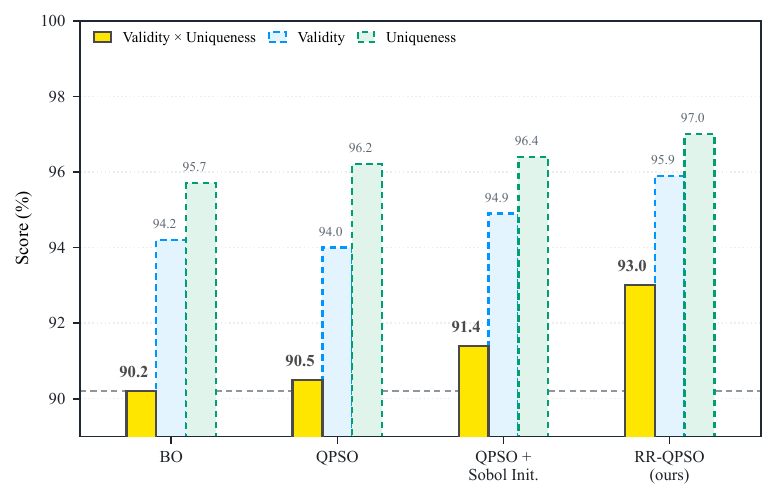}
  \caption{Optimizer performance comparison on the 9-heavy-atom QMG benchmark.
  The bars report validity, uniqueness, and their product. QPSO-based runs use
  $M=64$ particles, $T=150$ iterations, and 5000-shot molecular evaluation.
  RR-QPSO gives the highest validity--uniqueness product in this experimental
  setting.}
  \label{fig:performance}
\end{figure}

As shown in Fig.~\ref{fig:performance}, BO gives $V=94.2\%$, $U=95.7\%$, and $V\times U=90.2\%$. QPSO without Sobol initialization gives a similar product of $90.5\%$, mainly due to higher uniqueness. Adding Sobol initialization increases the product to $91.4\%$ in this single-seed run, suggesting that low-discrepancy initialization improves initial coverage but is not sufficient by itself. RR-QPSO gives the highest score under the same $M=64$, $T=150$ setting, with $V=95.9\%$, $U=97.0\%$, and $V\times U=93.0\%$. This trend suggests that the larger gain comes from the combination of rank-refined mean-best guidance and fitness-guided refinement. Since the comparison uses a fixed experimental protocol, the results should be viewed as benchmark-level evidence rather than a complete statistical ranking.

\subsection{Effect of Particle Count}
The second experiment studies the effect of the particle count $M$ in RR-QPSO. The optimizer design is fixed, and only $M$ is varied. We evaluate $M\in\{16,32,48,64,96,128\}$ with $T=150$ and 5000 shots per candidate.

\begin{figure}[!t]
  \centering
  \includegraphics[width=0.98\columnwidth]{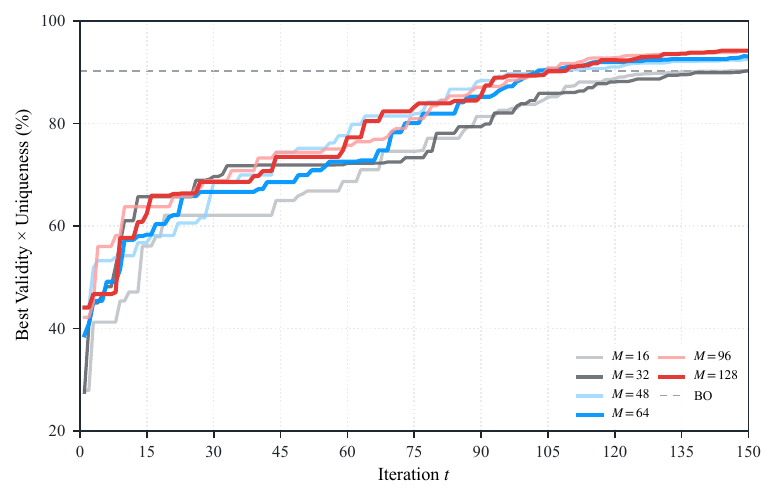}
  \caption{Effect of particle count on RR-QPSO convergence. Curves show the
  best validity--uniqueness product observed up to each iteration for different
  swarm sizes $M$. The dashed line indicates BO.}
  \label{fig:particle_count}
\end{figure}

Fig.~\ref{fig:particle_count} and Table~\ref{tab:particle_count} show that the final $V \times U$ generally improves as the particle count increases. Smaller swarms are competitive with BO, while larger swarms provide broader search coverage and converge to higher final scores. The best result is obtained at $M = 128$, where RR-QPSO reaches $V = 97.5\%$, $U = 96.6\%$, and $V \times U = 94.2\%$ $(0.9420)$. This is an absolute improvement of $0.040$, or $4.0$ percentage points, over BO, which reaches $V \times U = 0.9020$ under the same molecular-generation protocol.

Execution time depends on both $M$ and the multi-GPU worker schedule. Since particle evaluations are distributed across GPU workers, runtime is an implementation-dependent measurement rather than a strictly monotonic function of $M$. The particle-count study therefore focuses on optimization behavior, while detailed scaling efficiency would require a separate systems benchmark.

\begin{table}[!t]
\centering
\caption{Effect of particle count on validity, uniqueness, execution time in hours, and $V \times U$ for the full RR-QPSO configuration.}
\label{tab:particle_count}
\begin{tabular}{ccccc}
\toprule
\textbf{$M$} & \textbf{V (\%)} & \textbf{U (\%)} & \textbf{$V \times U$ (\%)} & \textbf{Time (h)} \\
\midrule
16  & 95.2 & 94.8 & 90.2 & 7.21 \\
32  & 94.4 & 95.6 & 90.2 & 15.44 \\
48  & 96.0 & 96.3 & 92.4 & 21.59 \\
64  & 95.9 & 97.0 & 93.0 & 47.12 \\
96  & 96.6 & 97.3 & 94.0 & 43.67 \\
\textbf{128} & \textbf{97.5} & \textbf{96.6} & \textbf{94.2} & \textbf{58.92} \\
\bottomrule
\end{tabular}
\end{table}

\begin{figure*}[!t]
  \centering
  \includegraphics[width=0.98\textwidth]{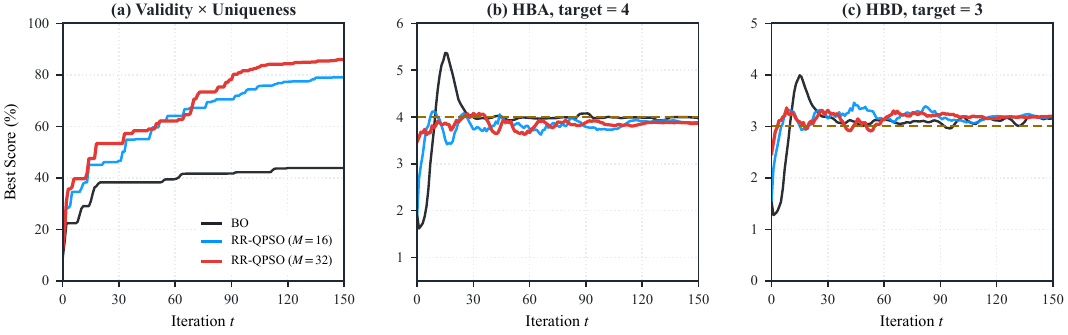}
  \caption{Multi-objective optimization targeting $\mathrm{HBA}=4$ and
  $\mathrm{HBD}=3$. BO and RR-QPSO are evaluated under the same scalarized
  target-property objective. RR-QPSO retains a higher validity--uniqueness
  product while keeping the generated molecules close to the target HBA and HBD
  region in this scalarized setting.}
  \label{fig:multi_objective}
\end{figure*}

\subsection{Multi-Objective Optimization}
The final experiment evaluates RR-QPSO in a property-guided setting. In addition to validity and uniqueness, the optimizer targets hydrogen-bond acceptor (HBA) and hydrogen-bond donor (HBD) counts, with desired values $\mathrm{HBA}=4$ and $\mathrm{HBD}=3$.

This setting introduces a trade-off between molecular quality and property matching. The optimizer must keep molecules valid and diverse while moving the generated distribution toward the target chemical region. We use a scalarized fitness function
\begin{equation}
F_{\mathrm{MO}}
=
(V\times U)
\left[
(1-\lambda)
+
\lambda C_{\mathrm{prop}}
\right],
\end{equation}
where $\lambda$ controls the property-guidance strength. The property closeness term is
\begin{equation}
C_{\mathrm{prop}}
=
\exp\left(
-\frac{1}{2}
\left[
\left(\frac{\bar{h}_{\mathrm{HBA}}-4}{\sigma_{\mathrm{HBA}}}\right)^2
+
\left(\frac{\bar{h}_{\mathrm{HBD}}-3}{\sigma_{\mathrm{HBD}}}\right)^2
\right]
\right),
\end{equation}
where $\bar{h}_{\mathrm{HBA}}$ and $\bar{h}_{\mathrm{HBD}}$ are the mean HBA and HBD counts of the generated valid molecules. In our experiments, we use $\lambda=0.40$ and $\sigma_{\mathrm{HBA}}=\sigma_{\mathrm{HBD}}=1$.

As shown in Fig.~\ref{fig:multi_objective}, both optimizers move the mean HBA and HBD values close to the target region. At the final iteration, BO reaches $V\times U=43.8\%$, $\overline{\mathrm{HBA}}=3.97$, and $\overline{\mathrm{HBD}}=3.16$, whereas RR-QPSO with $M = 32$ reaches $V \times U = 79.0\%$, $\overline{\mathrm{HBA}}=3.88$, and $\overline{\mathrm{HBD}}=3.15$. Thus, the main difference is not the final mean property value alone, but the ability to retain a higher validity--uniqueness product while staying near the target region. In this experiment, the population-based search maintains molecular quality more effectively under the added property constraint.

The current scalarization has a natural limitation: it uses mean HBA and HBD values, so it measures distribution-level centering rather than the full spread of generated properties. A fuller multi-objective analysis could also report property histograms, Pareto-front behavior, or target-hit ratios. Within the present benchmark, the result shows that RR-QPSO can incorporate property-guided objectives beyond the single $V\times U$ objective.

\section{Conclusion}
This paper studies the parameter-optimization bottleneck in QMG, where each
candidate circuit setting requires stochastic sampling, molecular decoding, and
chemical validity analysis. RR-QPSO is introduced as a population-based
alternative to BO, combining Sobol-based initialization, rank-refined mean-best
guidance, and validity--uniqueness-aware refinement. This design is well matched
to QMG because candidate evaluations are costly but mutually independent, making
them suitable for multi-GPU execution.

The results indicate that optimizer design can affect QMG output quality. On the
9-heavy-atom benchmark, RR-QPSO improves the validity--uniqueness product from
$0.9020$ for BO to $0.930$ at $M=64$, and reaches $0.9420$ at $M=128$. The
particle-count study suggests that larger swarms can improve search coverage,
although runtime depends on implementation-level scheduling. In the
multi-objective experiment, RR-QPSO maintains higher validity--uniqueness
performance than BO while keeping the mean HBA and HBD values near the target
region. Overall, the results support an optimizer-level route for improving QMG
without changing the chemistry-inspired circuit or molecular decoding pipeline.
Future work should include multi-seed evaluation, larger molecular settings,
longer optimization budgets, and more detailed property-distribution analysis
for property-guided molecular design.

\section*{Acknowledgment}
The authors gratefully acknowledge the National Center for High-performance Computing (NCHC), National Institutes of Applied Research (NIAR), Taiwan, for providing computational resources and research infrastructure. During manuscript preparation, the authors used OpenAI ChatGPT for language editing, grammar refinement, and wording suggestions. The authors reviewed and revised all AI-assisted text and take full responsibility for the final content.

\bibliographystyle{ieeetr}
\bibliography{reference}

@article{rebentrost2014quantum,
  title={Quantum support vector machine for big data classification},
  author={Rebentrost, Patrick and Mohseni, Masoud and Lloyd, Seth},
  journal={Physical review letters},
  volume={113},
  number={13},
  pages={130503},
  year={2014},
  publisher={APS}
}

@article{havlivcek2019supervised,
  title={Supervised learning with quantum-enhanced feature spaces},
  author={Havl{\'\i}{\v{c}}ek, Vojt{\v{e}}ch and C{\'o}rcoles, Antonio D and Temme, Kristan and Harrow, Aram W and Kandala, Abhinav and Chow, Jerry M and Gambetta, Jay M},
  journal={Nature},
  volume={567},
  number={7747},
  pages={209--212},
  year={2019},
  publisher={Nature Publishing Group UK London}
}

@article{chen2024validating,
  title={Validating Large-Scale Quantum Machine Learning: Efficient Simulation of Quantum Support Vector Machines Using Tensor Networks},
  author={Chen, Kuan-Cheng and Li, Tai-Yue and Wang, Yun-Yuan and See, Simon and Wang, Chun-Chieh and Wille, Robert and Chen, Nan-Yow and Yang, An-Cheng and Lin, Chun-Yu},
  journal={Machine Learning: Science and Technology},
  year={2024}
}

@article{cerezo2021variational,
  title={Variational quantum algorithms},
  author={Cerezo, Marco and Arrasmith, Andrew and Babbush, Ryan and Benjamin, Simon C and Endo, Suguru and Fujii, Keisuke and McClean, Jarrod R and Mitarai, Kosuke and Yuan, Xiao and Cincio, Lukasz and others},
  journal={Nature Reviews Physics},
  volume={3},
  number={9},
  pages={625--644},
  year={2021},
  publisher={Nature Publishing Group UK London}
}

@inproceedings{an2025quantum,
  title={Quantum Pointwise Convolution: A Flexible and Scalable Approach for Neural Network Enhancement},
  author={An Ning and Li, Tai-Yue and Chen, Nan-Yow},
  booktitle={2025 International Conference on Quantum Communications, Networking, and Computing (QCNC)},
  pages={371-378},
  year={2025},
  organization={IEEE}
}

@inproceedings{tai2022quantum,
  title={Classification of Tumor Metastasis Data by Using Quantum kernel-based Algorithms},
  author={Li, T-Y and Venugopala Reddy Mekala and Ng, K-L and Su, C-F},
  booktitle={2022 IEEE 22nd International Conference on Bioinformatics and Bioengineering (BIBE)},
  pages={351-354},
  year={2022},
  organization={IEEE}
}

@article{sam2026dual,
  title={Scalable Tensor Network Simulation for Quantum-Classical Dual Kernel},
  author={Sam, Mei Ian and Li, Tai-Yu},
  journal={arXiv preprint arXiv:2602.01330},
  year={2026}
}

@article{rai2026hybrid,
  title={Hybrid Classical-Quantum Neural Networks for Multi-Characteristic Co-Optimization of Recessed-Gate {AlGaN/GaN} {MIS-HEMTs}},
  author={Rai, Rushat and Chang, Pei-Jie and Nguyen, Doan Viet and Chiu, Yuan-Chieh and Tumilty, Niall and Wang, Yun-Yuan and See, Simon and Lee, Wen-Jay and Li, Tai-Yue and Chen, Nan-Yow and Wu, Tian-Li},
  journal={arXiv preprint arXiv:2605.27420},
  year={2026}
}

@article{wang2026mpmqir,
  title={{MPM-QIR}: Measurement-Probability Matching for Quantum Image Representation and Compression via Variational Quantum Circuit},
  author={Wang, Chong-Wei and Sam, Mei Ian and Kuo, Tzu-Ling and Chen, Nan-Yow and Li, Tai-Yue},
  journal={arXiv preprint arXiv:2601.03855},
  year={2026}
}

@article{sam2026grover,
  title={Iterative Matrix Product State Simulation for Scalable Grover's Algorithm},
  author={Sam, Mei Ian and Kuo, Tzu-Ling and Li, Tai-Yue},
  journal={arXiv preprint arXiv:2601.03832},
  year={2026}
}

@article{hsu2025qae,
  title={Quantum Adaptive Excitation Network with Variational Quantum Circuits for Channel Attention},
  author={Hsu, Yu-Chao and Chen, Kuan-Cheng and Li, Tai-Yue and Chen, Nan-Yow},
  journal={arXiv preprint arXiv:2507.11217},
  year={2025}
}

@article{polishchuk2013estimation,
  title={Estimation of the size of drug-like chemical space based on {GDB-17} data},
  author={Polishchuk, Pavel G and Madzhidov, Timur I and Varnek, Alexandre},
  journal={Journal of Computer-Aided Molecular Design},
  volume={27}, number={8}, pages={675--679}, year={2013}, publisher={Springer}
}

@article{gomezbombarelli2018automatic,
  title={Automatic chemical design using a data-driven continuous representation of molecules},
  author={G{\'o}mez-Bombarelli, Rafael and Wei, Jennifer N and Duvenaud, David and Hern{\'a}ndez-Lobato, Jos{\'e} Miguel and S{\'a}nchez-Lengeling, Benjam{\'\i}n and Sheberla, Dennis and Aguilera-Iparraguirre, Jorge and Hirzel, Timothy D and Adams, Ryan P and Aspuru-Guzik, Al{\'a}n},
  journal={ACS Central Science},
  volume={4}, number={2}, pages={268--276}, year={2018}, publisher={ACS Publications}
}

@inproceedings{jin2018junction,
  title={Junction tree variational autoencoder for molecular graph generation},
  author={Jin, Wengong and Barzilay, Regina and Jaakkola, Tommi},
  booktitle={International Conference on Machine Learning (ICML)},
  year={2018}
}

@article{segler2018generating,
  title={Generating focused molecule libraries for drug discovery with recurrent neural networks},
  author={Segler, Marwin H S and Kogej, Thierry and Tyrchan, Christian and Waller, Mark P},
  journal={ACS Central Science},
  volume={4}, number={1}, pages={120--131}, year={2018}, publisher={ACS Publications}
}

@article{decao2018molgan,
  title={{MolGAN}: An implicit generative model for small molecular graphs},
  author={De Cao, Nicola and Kipf, Thomas},
  journal={arXiv preprint arXiv:1805.11973},
  year={2018}
}

@article{biamonte2017quantum,
  title={Quantum machine learning},
  author={Biamonte, Jacob and Wittek, Peter and Pancotti, Nicola and Rebentrost, Patrick and Wiebe, Nathan and Lloyd, Seth},
  journal={Nature},
  volume={549}, number={7671}, pages={195--202}, year={2017}, publisher={Nature Publishing Group}
}

@article{dallaire2018quantum,
  title={Quantum generative adversarial networks},
  author={Dallaire-Demers, Pierre-Luc and Killoran, Nathan},
  journal={Physical Review A},
  volume={98},
  number={1},
  pages={012324},
  year={2018},
  publisher={APS}
}

@article{liu2018differentiable,
  title={Differentiable learning of quantum circuit {Born} machines},
  author={Liu, Jin-Guo and Wang, Lei},
  journal={Physical Review A},
  volume={98},
  number={6},
  pages={062324},
  year={2018},
  publisher={APS}
}

@article{li2021drug,
  title={Drug discovery approaches using quantum machine learning},
  author={Li, Junde and Alam, Mahabubul and Sha, Congzhou M and Wang, Jian and Dokholyan, Nikolay V and Ghosh, Swaroop},
  journal={arXiv preprint arXiv:2104.00746},
  year={2021}
}

@article{thomas2025qca,
  title={{QCA-MolGAN}: Quantum Circuit Associative Molecular {GAN} with Multi-Agent Reinforcement Learning},
  author={Thomas, Aaron Mark and Chen, Yu-Cheng and Valencia, Hubert Okadome and Jose, Sharu Theresa and Wu, Ronin},
  journal={arXiv preprint arXiv:2509.05051},
  year={2025}
}

@article{chen2025chemistry,
  title={Exploring Chemical Space with Chemistry-Inspired Dynamic Quantum Circuits in the NISQ Era},
  author={Chen, Lung-Yi and Li, Tai-Yue and Li, Yi-Pei and Chen, Nan-Yow and You, Fengqi},
  journal={Journal of Chemical Theory and Computation},
  year={2025},
  doi={10.1021/acs.jctc.5c00305}
}

@article{sun2012quantum,
  title={Quantum-behaved particle swarm optimization: analysis of individual particle behavior and parameter selection},
  author={Sun, Jun and Fang, Wei and Wu, Xiaojun and Palade, Vasile and Xu, Wenbo},
  journal={Evolutionary Computation},
  volume={20}, number={3}, pages={349--393}, year={2012}, publisher={MIT Press}
}

@inproceedings{sun2004particle,
  title={Particle swarm optimization with particles having quantum behavior},
  author={Sun, Jun and Feng, Bin and Xu, Wenbo},
  booktitle={Proceedings of the 2004 Congress on Evolutionary Computation},
  volume={1},
  pages={325--331},
  year={2004},
  organization={IEEE},
  doi={10.1109/CEC.2004.1330875}
}

@inproceedings{snoek2012practical,
  title={Practical Bayesian optimization of machine learning algorithms},
  author={Snoek, Jasper and Larochelle, Hugo and Adams, Ryan P.},
  booktitle={Advances in Neural Information Processing Systems},
  volume={25},
  pages={2951--2959},
  year={2012}
}

@article{sobol1967distribution,
  title={On the distribution of points in a cube and the approximate evaluation of integrals},
  author={Sobol', Ilya M.},
  journal={USSR Computational Mathematics and Mathematical Physics},
  volume={7},
  number={4},
  pages={86--112},
  year={1967},
  publisher={Elsevier}
}

@incollection{owen1995randomly,
  title={Randomly permuted $(t,m,s)$-nets and $(t,s)$-sequences},
  author={Owen, Art B.},
  booktitle={Monte Carlo and Quasi-Monte Carlo Methods in Scientific Computing},
  series={Lecture Notes in Statistics},
  volume={106},
  pages={299--317},
  year={1995},
  publisher={Springer},
  doi={10.1007/978-1-4612-2552-2_19}
}

\end{document}